\documentstyle[aps,epsfig,twocolumn]{revtex}
\def\om{\omega}
\def\Om{\Omega}
\def\al{\alpha}
\def\be{\begin{equation}}
\def\ee{\end{equation}}
\def\eea{\end{eqnarray}}
\def\bea{\begin{eqnarray}}
\newcommand{\br}{{\bf r}}

\title{Electronic properties of single and coupled anisotropic quantum dots\\
in a magnetic field, spin interactions and switching.}
\author{Alexander P.Itin\thanks{e-mail address: alx\_it@yahoo.com}}
\address{
Space Research Institute, Russian Academy of Sciences \\
 Profsoyuznaya str. 84/32, 117997 Moscow, Russia.
}
\date{\today}
\begin{document}
\maketitle
\begin{abstract}
We determined the eigenstates of a single electron in a parabolic
anisotropic 2D quantum dot in a magnetic field. Using obtained expressions
for these eigenstates, we study the spin coupling $J$ between two electrons
located in two laterally coupled anisotropic quantum dots (QD). The exhange
coupling $J$ is calculated using the Heitler-London and
Hund-Mulliken approaches. We found that the exhange $J$ changes sign at
certain values of parameters of the system, in particular at certain 
anisotropy of the QD. Therefore, we present a new method to switch on and off
the spin coupling between QD: switching by means of changing shape of the QD.
Switching the spin coupling is essential for quantum computation using
electronic spins as qubits. We note that our calculations can be applied to the 
system of vertically coupled QD as well.
\end{abstract}
%%%%%%%%%%%%%%%%%%%%%%%%%%%%%%%%%%%%%%%%%%%%%%%%%%%%%%%%%%%%%%%%%%%%%%%%%%%%

%\pacs{PACS numbers: 73.20.Dx Electron states in low-dimensional
%structures (superlattices, quantum well structures and multilayers)
%85.30.Vw Low-dimensional quantum devices
%(quantum dots, quantum wires etc.)
%03.67.Lx Quantum computation}

\section{Introduction.}
The last decade has seen a great interest in quantum dots, i.e. little islands
of electron gas confined by an artificial potential  (see Ref. \cite{int1}).
One of the most challenging proposals concerning QD is the idea of using the
electron spin in quantum dots as the basic information carrier (the qubit)
in a quantum computer (see, for example, Ref. \cite{Burc}). Quantum logic gates
between these qubits are effected 
by allowing the electrons to tunnel between two coupled quantum dots.
For the application of coupled quantum dots
as a quantum gate, it is important that the
coupling between the spins can be switched on and off via externally controlled
parameters such as magnetic field
and interdot distance. In the recent paper of G.Burcard et al.$^2 $ the 
spin interaction for two laterally coupled
{\em isotropic} QD defined in a two-dimensional electron system (2DES) was 
calculated as a function of these external parameters and it was shown that
 the interaction $J$ can be switched on and off with exponential sensivity
 to that parameters. In the present paper, we consider a different setup 
 consisting of two laterally coupled {\em
anisotropic} quantum dots with a magnetic field applied perpendicular to 
that 2DES (see Fig.1). We especially interested in calculation of interaction 
$J$ as a function of anisotropy of the QD since our proposal is to switch on 
and off the interaction $J$ by changing shape of the QD, e.g. by applying 
gate voltage to each quantum dot. We note that in the
setup being proposed in Ref. \cite{Burc} it was assumed that the interdot 
distance is
controlled by varying the barrier height between the QD. In real experimental
setup varying the barrier height might lead to changing shape of the QD
from circular to elliptic. Also, probably in future quantum computers based on
large integrated circuts of QD it will be simplier to change shape of the QD 
(and, in case of neceserrity, interdot distance), than change interdot distance
alone, preserving circular shape of the QD.
This is one of the reasons which motivated our study. 

In order to study coupled anisotropic quantum dots, one needs to find firstly
the eigenstates of a single electron in a single anisotropic quantum dot
in a magnetic field.
The latter problem is of fundamental interest in mesoscopic physics. 
Surprisingly, it seems that this important problem was not solved so far.

The eigenstates of a single two-dimensional electron confined by a radial
potential of the form $\frac{1}{2}m_e \om^2 r^2$ with a magnetic field 
perpendicular to the plane of the system are called Fock-Darwin states.  
That system was investigated by Fock$^3$  and Darwin$^4$ a long time ago.
The system where an electron is confined by an anisotropic potential
of the form $\frac{1}{2}m_e (\om_x^2 x^2 + \om_y^2 y^2)$ was studied only
recently by A.V. Madhav and Tapash Chakraborty.$^5$ Their studies
were motivated, in particular, by the experiments on anisotropic quantum dots
(see Ref. \cite{exp}).  
They obtained energy eigenvalues of the system, but not the eigenfunctions
of it. In their paper, the transformation of coordinates and momenta of the
system was performed. In the new Hamiltonian, variables are separated and one 
can easily obtain both the energy eigenvalues and the eigenstates of the system.
However, in order to obtain the eigenstates of the system in old (initial) 
coordinates, one needs to transform the eigenfunctions of the new Hamiltonian.
 It was not done in Ref. \cite{MC}. In the section II of the present 
 paper, we performed such a transformation and obtained exact analytical
expressions for the eigenstates of that system. Note that the system of
an anisotropic quantum dot in a magnetic field was considered also in Ref.
\cite{seelig}, where only {\em approximate} solution for the ground state
of the system was obtained (see Appendix D in that paper). 
In the section III we study
coupled anisotropic QD using expressions obtained in section II.
Following Ref. \cite{Burc}, we employ Heitler-London approximation and 
Hund-Mulliken approach. In section IV we give brief conclusion.

\section{Single anisotropic quantum dot.}
Let us consider a single electron in a lateral anisotropic parabolic
confinement potential in the presence of a perpendicular magnetic field.
The Hamiltonian is
\begin{equation}
{\cal{H}} = \frac{1}{2m_e} \left[ p-\frac{e}{c}A \right]^2+V_{conf}(x,y),
\end{equation}
where $m_e$ is the effective mass of the electron and the confinement potential 
is
\begin{equation}
V_{conf}(x,y)=\frac{1}{2}m_e(\om_x^2 x^2+ \om_y^2 y^2).
\end{equation}
Choosing the symmetric gauge vector potential  $A=\frac{1}{2}B(-y,x,0)$,
we get
\begin{eqnarray}
{\cal{H}} &=& \frac{1}{2m_e}[p_x^2+\Om_1^2 x^2 + p_y^2 + \Om_2^2 y^2 +
 m_e \om_c(y p_x-xp_y)], \nonumber\\
\Om_{1,2}^2 &=&m_e^2(\om_{x,y}^2+\frac{1}{4}\om_c^2), \label{openh}\\
 \om_c&=&eB/m_e c.\nonumber
\end{eqnarray}

Following Ref. \cite{MC}, we make the following transformations:
\begin{eqnarray}
x &=& \mu q_1 + \eta p_2, \nonumber\\
y &=& \mu q_2 + \eta p_1, \label{trans}\\
p_x &=& \mu p_1 - \zeta q_2, \nonumber\\
p_y &=& \mu p_2 - \zeta q_1, \nonumber
\end{eqnarray}
where 
\begin{eqnarray}
\mu&=&\frac{\sqrt{\Om_3^2+\Om_1^2-\Om_2^2}}{\Om_3\sqrt{2}}, 
\eta=\frac{\sqrt{\Om_3^2+\Om_2^2-\Om_1^2}}{\Om_3\sqrt{\Om_1^2+\Om_2^2}},
\nonumber\\
\zeta&=&\frac{\sqrt{(\Om_3^2+\Om_2^2-\Om_1^2)(\Om_1^2+\Om_2^2)}}{2\Om_3},\\
\Om_3^2&=&[(\Om_1^2-\Om_2^2)^2+2m_e^2 \om_c^2(\Om_1^2+\Om_2^2)]^{1/2}.
\nonumber
\end{eqnarray}

The transformations (\ref{trans}) are consistent with the commutation relations 
$[p_i,q_j]=-i \hbar \delta_{ij}, [q_i,q_j]=0$ and the resulting Hamiltonian is
diagonal (see also Ref. \cite{MC}) :
\begin{equation}
H=\frac{1}{2m}(\al_1^2 p_1^2+\al_2^2 p_2^2 + \beta_1^2q_1^2+\beta_2^2q_2^2),
 \end{equation}
 where
\begin{eqnarray}
\al_1^2=\frac{\Om_1^2+3\Om_2^2+\Om_3^2}{2(\Om_1^2+\Om_2^2)},
\beta_1^2=\frac{1}{4}(3\Om_1^2+\Om_2^2+\Om_3^2), \\
\al_2^2=\frac{3\Om_1^2+\Om_2^2-\Om_3^2}{2(\Om_1^2+\Om_2^2)},
\beta_2^2=\frac{1}{4}(\Om_1^2+3\Om_2^2-\Om_3^2).\nonumber \\
\end{eqnarray}
One can easily obtain eigenvalues and eigenstates of the Hamiltonian:
\begin{eqnarray}
E_{m,n}&=&(m+\frac{1}{2})\hbar \om_1+(n+\frac{1}{2})\hbar \om_2, \nonumber \\
\psi_{m,n}(q_1,q_2)&=&c_{m,n}e^{-\frac{\gamma_1^2 q_1^2}{2}}
e^{-\frac{\gamma_2^2 q_2^2}{2}}
{\rm H}_m(\gamma_1 q_1){\rm H}_n(\gamma_2 q_2),\label{psi}
\end{eqnarray}
where $ c_{m,n}=\left(\frac{\gamma_1 \gamma_2 }{\pi 2^{m+n}m!n!}\right)^{1/2}$,
 $\om_i=\al_i \beta_i/m_e$,
 $\gamma_i^2=\beta_i/(\al_i \hbar)$, ${\rm H_k} $ is the Hermit polynomial.
The question is how one can return back to coordinates (x,y), i.e. how
to obtain eigenstates $\Psi_{m,n}(x,y)$ of the Hamiltonian (\ref{openh}).

Since $ \psi(q_1,q_2) \equiv \langle q_1,q_2| \Psi\rangle$,
 $ \Psi(x,y) \equiv \langle x,y| \Psi\rangle$, and 
 $|\Psi\rangle =\int_{-\infty}^{+\infty} dq_1 dq_2 |q_1,q_2\rangle 
 \psi(q_1, q_2), $
we have the following formula for the eigenstates $\Psi_{m,n}(x,y)$:
\begin{equation}
\Psi_{m,n}(x,y)= \int_{-\infty}^{+\infty} dq_1 dq_2 
\langle x,y|q_1,q_2 \rangle \psi_{m,n}(q_1, q_2) \label{psixy}
\end{equation}
Let us denote $\Phi (x,y,q_1,q_2) = \langle x,y|q_1,q_2 \rangle $.
Function  $\Phi (x,y,q_1,q_2)$ is the eigenstate of operators
$ \hat{q}_1, \hat{q}_2 $ in the (x,y) representation 
and should obey the following relations:

 \begin{eqnarray}
 \hat{q}_1\Phi =[\mu\hat{x}- \eta \hat{p}_y]\Phi&=&q_1 \Phi, \\
 \hat{q}_2\Phi=[\mu\hat{y}- \eta\hat{p}_x]\Phi&=&q_2 \Phi. \nonumber
 \end{eqnarray}
On the other hand, $\Phi^* (x,y,q_1,q_2)=\langle q_1,q_2|x,y \rangle$
('*' denotes complex conjugating) and therefore it should obey the following
 relations:
 \begin{eqnarray}
 \hat{x}\Phi^* =[ \mu\hat{q}_1+
 \eta\hat{p}_2]\Phi^* &=&x \Phi^*, \\
 \hat{y}\Phi^*=[\mu\hat{q}_2+
 \eta\hat{p}_2]\Phi^*&=&y \Phi^*. \nonumber
 \end{eqnarray}

Therefore, we finally have the following system of differential equations
 defining $\Phi (x,y,q_1,q_2)$:
$$
\left\{
\begin{array}{lcl}
(i\hbar \eta\frac{ \partial}{\partial y} +\mu x)\Phi (x,y,q_1,q_2) &=&
q_1 \Phi (x,y,q_1,q_2),\nonumber\\ 
(i\hbar \eta\frac{ \partial}{\partial x} +\mu y)\Phi (x,y,q_1,q_2) &=&
q_2 \Phi (x,y,q_1,q_2), \\ 
(-i\hbar \eta\frac{ \partial}{\partial q_2} +\mu q_1)\Phi^* (x,y,q_1,q_2)&=&
x \Phi^* (x,y,q_1,q_2),\nonumber\\
(-i\hbar \eta\frac{ \partial}{\partial q_1} +\mu q_2) \Phi^* (x,y,q_1,q_2)&=&
y \Phi^* (x,y,q_1,q_2).\nonumber
\end{array}
\right.
$$

One can find that solution of this system has the following form:
\begin{equation}
\Phi (x,y,q_1,q_2)=C {\rm exp}\left\{\frac{i}{\eta \hbar}(\mu xy+ \mu q_1 q_2-
 q_1y-q_2x)\right\},
\label{psiq}
\end{equation}
where C is a constant.
From (\ref{psiq},\ref{psixy},\ref{psi}) we have

\begin{equation}
\Psi_{m,n}(x,y)= 
\hat{c}_{m,n} e^{\frac{i \mu xy}{\eta \hbar}} I_{m,n}(\gamma_1 \nu x, 
\gamma_2 \nu y), 
\label{psimn}
\end{equation}
where $\hat{c}_{m,n}=Cc_{m,n}/{\gamma_1 \gamma_2 }$, 
$\nu=(\mu ^2+\eta^2 \hbar^2
 \gamma_1^2 \gamma_2^2)^{-1/2}, and$

\begin{eqnarray}
I_{m,n}(z_1,z_2) &=& \nonumber\\
\int\limits_{-\infty}^{+\infty}
\psi_{m,n}\left( \frac{q_1}{\gamma_1}, \frac{q_2}{\gamma_2} \right)
&e&^{i(\sigma q_1 q_2-q_2 z_1 d-q_1 z_2d)} dq_1 dq_2,
\end{eqnarray}
where $\sigma=\mu/(\eta \hbar \gamma_1 \gamma_2 )$, $d=\sqrt{1+\sigma^2}$.
Let us firstly find the ground state of our system $\Psi_{00}(x,y).$
Diffirentiating expression for $I_{00}(z_1,z_2)$ over $z_1$ and $z_2$ and
solving obtained diffirential equations, one can find that
\begin{equation}
I_{00}(z_1,z_2)=\hat{C}_{00} e^{-\frac{z_1^2}{2}-\frac{z_2^2}{2}}
e^{-i \sigma z_1 z_2},\label{i00}
\end{equation}
where $\hat{C}_{00}$ is a constant.
Therefore, the ground state of the system is
\begin{equation}
\Psi_{00}(x,y)=C_{00}e^{ \frac{i \mu xy}{\eta \hbar}(1-\nu^2)}
 e^{-\frac{\kappa_x^2  x^2 }{2}}
 e^{-\frac{\kappa_y^2  y^2 }{2}}  ,\label{psi00}
\end{equation}
where normalization constant is
$ C_{00}= \sqrt{ \frac{\kappa_x \kappa_y }{\pi}} $, $\kappa_x= \nu \gamma_1$,
$\kappa_y= \nu \gamma_2$.
One can see that $ C \hat{C}_{00}= \kappa_x \kappa_y/ \nu $.
Let us now note that ${\rm H}_n(q)=(2q)^n-\frac{n(n-1)}{1}(2q)^{n-2}+
\frac{n(n-1)(n-2)(n-3)}{1 \cdot 2} (2q)^{n-4}-..$

Let $L_k(z_1,z_2)= 
\int_{-\infty}^{+\infty}
(2q_2)^k F(q_1,q_2,z_2) e^{-i q_2 z_1 d} dq_1 dq_2$ ,
where $F$ is some function. One can see that
\begin{equation}
L_k(z_1,z_2)=\frac{2i}{d}\frac{\partial L_{k-1}}{\partial z_1}=...=
\left(\frac{2i}{d} \right)^k \frac{\partial^k L_0}{\partial z_1^k}.
\end{equation}
A similar procedure for the coordinate $z_2$ can be executed.
Now one can easily find
the following expression for $I_{m,n}(z_1,z_2):$ 

\begin{equation}
I_{m,n}(z_1,z_2)=\hat{D}^m_{z_2}\hat{D}^n_{z_1}I_{00}(z_1,z_2),\label{imn}
\end{equation}
where operators $\hat{D}^k_{\alpha}$ are defined as

\begin{eqnarray}
&\hat{D}^k_{\alpha}&={\rm H}_k \left(\frac{i}{d}\frac{\partial}{\partial 
\alpha} \right)=\\
&=&\left(\frac{2i}{d} \right)^k \frac{\partial^k}{\partial \alpha^k}-
\frac{k(k-1)}{1}\left(\frac{2i}{d} \right)^{k-2} 
\frac{\partial^{k-2}}{\partial \alpha^{k-2}}+...\nonumber
\end{eqnarray}
So, we finally have the following expression for the 
 eigenstates of the Hamiltonian (\ref{openh}):
\begin{equation}
\Psi_{m,n}(x,y)= \sqrt{\frac{\kappa_x \kappa_y}{\pi 2^{m+n}m!n!}} 
e^{\frac{i \mu xy}{\eta \hbar}} 
\hat{D}^m_{\kappa_y y}\hat{D}^n_{\kappa_x x} I(x,y),\label{final}
\end{equation}
where $I(x,y)= {\rm exp} \{-\frac{\kappa_x^2 x^2}{2}-\frac{\kappa_y^2 y^2}{2}-
i \sigma \kappa_x \kappa_y xy\}$.
The lowest excited states of the system are
\begin{eqnarray}
\Psi_{01}(x,y) &=& \frac{1}{d}\sqrt{\frac{2 \kappa_x \kappa_y}{\pi}}
 (\sigma \kappa_y y - i \kappa_x x )
 F_{00} (x,y), \nonumber\\
\Psi_{10}(x,y) &=& \frac{1}{d}\sqrt{\frac{2 \kappa_x \kappa_y}{\pi}}
(\sigma \kappa_x x - i \kappa_y y )
  F_{00} (x,y), \label{psi011}\\
\Psi_{11}(x,y) &=& \frac{2}{d^2}\sqrt{\frac{\kappa_x \kappa_y}{\pi}}                                        
[  \kappa_x \kappa_y xy(1-\sigma^2)+  \nonumber\\
&+&i \sigma (\kappa_x^2 x^2+\kappa_y^2 y^2) -i \sigma]
F_{00} (x,y),\nonumber
\end{eqnarray}
where $F_{00} (x,y)= e^{ \frac{i \mu xy}{\eta \hbar}(1-\nu^2)}
 e^{-\frac{\kappa_x^2 x^2}{2}
 -\frac{\kappa_y^2 y^2}{2}}=C_{00}^{-1} \Psi_{00}(x,y)$. 
Note that in degenerate case of isotropic dot we have $\Om_1=\Om_2$, $\nu=
\sigma=1$, 
and the eigenstates (\ref{psi011},\ref{psi00}) coincide with the corresponding
Fock-Darwin states.

\section{Coupled anisotropic quantum dots.}
\subsection{Model.}
The Hamiltonian which we use for the description of
 two laterally coupled quantum dots is
\begin{eqnarray}
H &=& \sum_{i=1,2} h({\bf r}_i,{\bf p}_i)+C,
\nonumber\\
h({\bf r},{\bf p}) &=& \frac{1}{2m_e}\left({\bf p}-\frac{e}{c}{\bf A}({\bf r})
\right)^2 +V(\bf r),\label{hamiltonian}\\
C&=&\frac{e^2}{\kappa\left| {\bf r}_1-{\bf r}_2\right|},\nonumber
\end{eqnarray}
where $C$ is the Coulomb interaction and $h$ is the
single-particle Hamiltonian. The dielectric constant $\kappa$ and the
effective mass $m_e$ are material parameters. For the lateral 
confinement V we choose the quartic potential 
\begin{equation}\label{potential}
V(x,y)=\frac{m_e\omega_x^2}{8 a^2}\left(x^2-a^2
\right)^2+ \frac{m_e\omega_y^2}{2} y^2, 
\end{equation}
which separates (for $x$ around $\pm a$) into two anisotropic
 harmonic wells in the limit
of large inter-dot distance, i.e. for $2a\gg 2 a_{Bx,y} $
, where $a$ is
half the distance between the centers of the dots, and the lateral effective 
Bohr radii $a_{Bx,y}= \sqrt{\hbar/m_e\omega_{x,y}}$ are a measure for the 
lateral extension of the electron wave function in the dots (see Fig. 1).
In Ref. \cite{Burc} the potential similar to (\ref{potential}) was considered,
but there $\omega_x=\omega_y=\omega_0$. It was also shown in Ref. \cite{Burc}
that spin-orbit contribution and Zeeman splitting can be neglected in relevant 
cases. For material parameters, we choose that of GaAs ($\kappa=13.1$, 
$m_e=0.067m$, where m is the mass of the electron).

\subsection{Heitler-London approach.}
Here we consider firstly the Heitler-London approximation, and than refine 
this approach by including (see subsection C) double occupancy 
in a Hund-Mulliken approach.
In the Heitler-London approach, one starts from
single-dot ground-state coordinate wavefunctions $\varphi(\bf r)$ and combines
them into the (anti-) symmetric  two-particle state vector
\begin{equation}
|\Psi_{\pm}\rangle = \frac{|12\rangle
\pm |21\rangle}{\sqrt{2(1\pm S^2)}},
\end{equation}
the positive (negative) sign
corresponding to the spin singlet (triplet) state,
and $S=\int d^2r\varphi_{+a}^{*}(\bf r)\varphi_{-a}(\bf r)=\langle 2|1\rangle$
denoting the overlap of the right and left orbitals.
A non-vanishing overlap implies that the electrons tunnel
between the dots (see also Ref. \cite{Burc}).
Here, $\varphi_{ -a}(\bf r)=\langle \bf r| 1\rangle$ and $\varphi_{+a}(\bf r)
=\langle \bf r|2\rangle$ denote the one-particle orbitals
centered at $\bf r=(\mp a,0)$, and $|ij\rangle = |i\rangle |j\rangle$ are
two-particle product states. The exchange energy is then obtained
as $J = \epsilon_{\rm t}-\epsilon_{\rm s} = 
\langle\Psi_{-}|H|\Psi_{-}\rangle -
\langle\Psi_{+}|H|\Psi_{+}\rangle$.
The single-dot orbital for anisotropic harmonic
confinement in two dimensions in a perpendicular magnetic field is the
ground state being obtained in section II. The ground state centered
at the origin is
\be
\varphi (x,y)=\sqrt{\frac{\kappa_x \kappa_y}{\pi}} e^{i\gamma xy} e^{-\kappa_x^2
x^2/2} e^{-\kappa_y^2 y^2/2},
\ee
where $\gamma=\mu(1-\nu^2)
/ \eta \hbar$, $\kappa_x$ and $\kappa_y$ were introduced in section II.

Shifting the single particle orbitals to $(\pm a,0)$ in the
presence of a magnetic field we obtain $\varphi_{\pm a}(x,y) = \exp(\pm
iya/2l_B^2)\varphi (x\mp a,y)$. The phase factor involving the magnetic
length $l_B=\sqrt{\hbar c/eB}$ is due to the gauge transformation ${\bf
A}_{\pm a}=B(-y,x\mp a,0)/2\rightarrow {\bf A}=B(-y,x,0)/2$ (see also 
Ref. \cite{Burc}). 
We rewrite the Hamiltonian, adding and subtracting the potential
of the single left (right) dot for electron 1(2) in $H$.
The Hamiltonian then takes the form $H=h_{-a}^0({\bf r}_1)+
h_{+a}^0({\bf r}_2)+W+C $, where $h_{ \pm a}^0({\bf r}_i)=({\bf
p}_i-e{\bf A}({\bf r}_i)/c)^2/2m_e + m_e\omega_x^2(x_i\mp a)^2/2+ 
 m_e\omega_y^2 y_i^2/2$, and $W({\bf r_1,r_2})=\sum_{i=1,2}V(x_i,y_i)-
 m_e\omega_x^2((x_1+a)^2+(x_2-a)^2)/2-
m_e\omega_y^2(y_1^2+y_2^2)/2$. The formal expression for $J$ is 
now
\begin{equation}\label{Jformal}
J = \frac{2S^2}{1-S^4}\left(\langle 12|C+W|12\rangle-\frac{{\rm Re}
\langle 12|C+W|21\rangle}{S^2}\right),
\end{equation}
where the overlap becomes $S= {\rm exp} (-\kappa_x^2 a^2-\kappa_y^{-2}
 l_{\gamma}^{-2})$.
Here we denoted $l_{\gamma}=(a\gamma-a/2l_B^2)^{-1}$.
Evaluation of the matrix elements of $C$ and $W$ provides us with 
the result,
\bea
J&=& \biggl[
c \kappa_y e^{-\kappa_x^2 a^2}
 {\rm K} \left( 1-\kappa_y^2 \kappa_x^{-2} \right)
 {\rm I_0 }(\kappa_x^2 a^2) \nonumber\\
&-&c\kappa_x e^{\kappa_y^{-2} l_{\gamma}^{-2}}
{\rm K} \left(1-\kappa_x^2 \kappa_y^{-2} \right)
 {\rm I_0} (\kappa_y^{-2} l_{\gamma}^{-2})
\nonumber\\
&+&\frac{3}{4} \frac{m_e \omega_x^2}{\kappa_x^2}(\kappa_x^2a^2+1) \biggr]/
\sinh(2\kappa_x^2 a^2+2\kappa_y^{-2} l_{\gamma}^{-2} ),\label{Jreal}
\eea
where $c=\sqrt{2/\pi}e^2/\kappa$, 
${\rm K}$ is the complete elliptic integral of the first kind, and ${\rm I_0}$
is 
the zeroth order Bessel function.
The first and second term in Eq.(\ref{Jreal}) are due to Coulomb interaction
$C$, where the exchange term enters with a minus sign. The last term comes 
from the confinement
potential $W$. Note that in the degenerate case of isotropic QD our result
coincide with that being obtained in Ref. \cite{Burc}.

We plotted $J$ as a function
of $\om_y$ and $\om_x$ at different values of magnetic field $B$ in Fig. 2.
Note singlet-triplet crossing ($J=0$) at certain parameters of the system.

\subsection{Hund-Mulliken approach.}
Following Ref. \cite{Burc}, we now apply the Hund-Mulliken approach to 
calculate the exchange energy of the double-dot system.
We introduce the orthonormalized
one-particle wave functions $\Phi _{\pm a}=(\varphi_{\pm }-
g\varphi_{\mp a})/\sqrt{1-2Sg+g^2}$, where $S$ again denotes the
 overlap of 
$\varphi_{-a}$ with $\varphi_{+a}$ and $g=(1-\sqrt{1-S^2})/S$. Using
 $\Phi _{\pm a}$, we generate four basis functions to which we 
diagonalize the two-particle Hamiltonian $H$: the states with
 double occupation, $\Psi^{\rm d}_{\pm a}({\bf r_1},{\bf r_2} )=
 \Phi_{\pm a}(\br_1)\Phi_{\pm a}(\br_2)$ and the states with single occupation,
 $\Psi^{\rm s}_{\pm}(\br_1,\br_2)=[\Phi_{+a}(\br_1)\Phi_{-a}(\br_2)\pm
\Phi_{-a}(\br_1)\Phi_{+a}(\br_2)]/\sqrt{2}$.
Diagonalization of
\begin{equation}\label{matrix}
H = 2\epsilon + \left(\begin{array}{cccc}
U&X&-\sqrt{2}t_{\rm H}&0\\
X&U&-\sqrt{2}t_{\rm H}&0\\
-\sqrt{2}t_{\rm H}&-\sqrt{2}t_{\rm H}&V_+&0\\
0&0&0&V_-
\end{array}\right)
\end{equation}
yields the eigenvalues
$\epsilon_{{\rm s}\pm} = 2\epsilon + U_{\rm H}/2+V_+\pm\sqrt{U_{\rm
H}^2/4+4t_{\rm H}^2}$, $\epsilon_{{\rm s} 0} = 2\epsilon + U_{\rm H} - 2X
+ V_+$ (singlet), and $\epsilon_{\rm t} = 2\epsilon + V_-$ (triplet),
where the following quantities were introduced :
\begin{eqnarray}
\epsilon &=& \langle\Phi_{\pm a}|h|\Phi_{\pm a}\rangle , \nonumber\\
t_{\rm H} &=& t- w = -\langle\Phi_{\pm
a}|h|\Phi_{\mp a}\rangle -
\langle\Psi^{\rm s}_+|C|
\Psi^{\rm d}_{\pm a}\rangle/\sqrt{2}, \nonumber\\
V &=& V_- - V_+ = \langle\Psi^{\rm
s}_-|C|\Psi^{\rm s}_-\rangle-
\langle\Psi^{\rm s}_+|C|\Psi^{\rm s}_+\rangle, \label{matrixelements} \\ 
X &=& \langle \Psi^{\rm d}_{\pm a}|C|\Psi^{\rm d}_{\mp a}\rangle, \nonumber\\
U_{\rm H} &=& U-V_+ +X \nonumber\\
&=& \langle\Psi^{\rm d}_{\pm a}
|C|\Psi^{\rm d}_{\pm a}\rangle-\langle\Psi^{\rm s}_+|C|\Psi^{\rm
s}_+\rangle+\langle\Psi^{\rm d}_{\pm a}|C|\Psi^{\rm d}_{\mp a}\rangle,\nonumber
\end{eqnarray}
The exchange energy is the gap
between the lowest singlet and the triplet state
\begin{equation}
J = \epsilon_{\rm t}-\epsilon_{{\rm s}-}=V - \frac{U_{\rm H}}{2} +
\frac{1}{2}\sqrt{U_{\rm H}^2 + 16t_{\rm H}^2}.
\label{HMresult}
\end{equation}
The results of evaluation of matrix elements are given in Appendix.
We plotted resulted $J$ as a function of 
 $\om_y$ and $\om_x$ at different values of magnetic field $B$ in Fig. 3.

\section{Concluding remarks.}
In summary, we obtained the eigenstates of two-dimensional anisotropic 
oscillator in a magnetic field (Eq. (\ref{final})) which in degenerate 
isotropic case become
the well-known Fock-Darwin states. We used obtained analytical expressions
to study a system of two coupled elliptic QD in a magnetic field.
We have calculated the exchange energy $J(B, a, \omega_x, \omega_y)$ between 
spins of coupled QD as a function of magnetic field, interdot distance
and parameters $\omega_x, \omega_y$ defining shape of a single quantum dot. 
We have shown that by varying $\omega_x$ or $ \omega_y$ 
the exchange coupling $J$ can be switched on and off.
This opens up the alternative way (compared to that being investigated in
Ref. \cite{Burc}) of performing quantum computing operations using coupled
QD as quantum gates. In the present paper, we didn't consider spin-orbit
contribution, Zeeman splitting and interplay between nonadiabaticity of
varying the system parameters and excitation of higher energy levels.
We consider these questions in a separate paper (see Ref. \cite{Itin}).
We emphasize that our results can easily be applied to 
the system of two {\em vertically} coupled three-dimensional QD with in-plane
magnetic field
$B$ being studied in Ref. \cite{seelig}, since the Hamiltonian of that problem
separates on $B$-independent part, which is merely a harmonic oscillator,
and $B$-dependent part which actually coincides with the Hamiltonian of two
{\em laterally} coupled anisotropic QD being studied in the present paper.  

To conclude, we believe that our results bring closer the day the first 
quantum computer$^{9-20}$ will begin to work.

\acknowledgments
The author wishes to thank Marat Khalili for illuminating discussions.
This work was supported in part by grant RFBR 00-01-00538. 
\appendix
\section{Hund-Mulliken matrix elements}\label{appendix}

Here, we give the explicit expressions for the matrix elements
defined in Eqs.~(\ref{matrix}) and (\ref{matrixelements}). The 
single-particle matrix elements are given by
\begin{eqnarray}
\epsilon = \frac{3}{32}\frac{m_e \om_x^2}{\kappa_x^4 a^2}
          & +& \frac{3}{8} \frac{S^2}{1 - S^2} 
           \frac{m_e \om_x^2}{\kappa_x^2}
             \left(1 + \kappa_x^2 a^2\right) +  \nonumber\\
             \frac{\hbar \alpha_1 \beta_1}{2 m_e} &+&
             \frac{\hbar \alpha_2 \beta_2}{2 m_e}, \\
t &=&  \frac{3}{8} \frac{S}{1 - S^2}
           \frac{m_e \om_x^2}{\kappa_x^2}
            \left(1 + \kappa_x^2  a^2\right),
\end{eqnarray}
where $S= {\rm exp} (-\kappa_x^2 a^2-\kappa_y^{-2}
 l_{\gamma}^{-2})$.

The (two-particle) Coulomb matrix elements 
are formally equal to that given in Ref. \cite{Burc},
where only $F_i$ have to be changed. We give here the
complete set of expressions for convenience:
\begin{eqnarray}
V_+ &=&c N^4 \left( 4 g^2 (1 + S^2) F_1 + (1 + g^2)^2 F_2 \right.\nonumber\\
                      &+& \left. 4 g^2 F_3 -  16 g^2 F_4\right),\\
V_- &=&c N^4 (1-g^2)^2 (F_2  -  S^2 F_3), \\
U   &=&c N^4 \left( (1 + g^4 + 2 g^2 S^2)F_1 + 2 g^2 F_2\right.\nonumber\\
                   &+& \left.2 g^2 S^2 F_3 - 8 g^2 F_4 \right),\\
X   &=&c N^4 \left[\left( (1 + g^4) S^2 + 2 g^2 \right) F_1 + 2 g^2 F_2 
\right.\nonumber\\
                      &+& \left.2 g^2 S^2 F_3 - 8 g^2 F_4 \right],\\
w   &=&c N^4 \left( - g (1 + g^2) (1 + S^2) F_1 - g (1 + g^2) F_2 \right.
  \nonumber\\
               &-& \left. g (1 + g^2) S^2 F_3 + (1 + 6 g^2 + g^4) S F_4 \right),
\end{eqnarray}
with $N=1/\sqrt{1-2Sg+g^2}$, $c= \frac{e^2}{\kappa} \sqrt{\frac{2}{\pi}}$,
and $g=(1-\sqrt{1-S^2})/S$.
Here, we make use of the functions
\begin{eqnarray}
F_1 &=& \kappa_x {\rm K} \left( 1-\kappa_x^2 \kappa_y^{-2} \right), \\
F_2 &=&  \kappa_y {\rm K} \left(1-\kappa_y^2 \kappa_x^{-2} \right)
   e^{-\kappa_x^2 a^2} {\rm I}_0\left(\kappa_x^2 a^2\right),\\
F_3 &=& \kappa_x {\rm K} \left( 1-\kappa_x^2 \kappa_y^{-2} \right)
e^{\kappa_y^{-2} l_{\gamma}^{-2}}
 {\rm I}_0\left(\kappa_y^{-2} l_{\gamma}^{-2}\right) ,\\
F_4 &=& \kappa_x {\rm K} \left( 1-\kappa_x^2 \kappa_y^{-2} \right)
 e^{\frac{\kappa_y^{-2} l_{\gamma}^{-2}-\kappa_x^2 a^2}{4}} \times
 \nonumber\\
   &\times& \sum_{k=-\infty}^{\infty}(-1)^k
   {\rm I}_{2k} \left( \frac{\kappa_x^2 a^2 + \kappa_y^{-2} 
l_{\gamma}^{-2}}{4}\right)
         {\rm I}_{2k}\left(i\frac{\kappa_x a}{2 \kappa_y 
l_{\gamma}}\right),\nonumber\\
\end{eqnarray}
where ${\rm I}_n$ denotes the Bessel function of $n$-th order.

\pagebreak
{\Large {\bf Figure captions.}}

\bigskip
\bigskip
\noindent Figure 1. Two coupled anisotropic QD with one valence electron per dot. Each electron is confined to the xy plane.
The magnetic field $B$ is perpendicular to the plane,
i.e. along the z axis. The quartic potential $V(x,y)$ is
given in Eq. (\ref{potential}) and is used to model the
coupling of two anisotropic harmonic wells centered at
$(\pm a,0,0)$. 
\bigskip

\noindent Figure 2. Exchange energy $J$ in units of meV 
plotted against anisotropy $\om_y/ \om_x$, as obtained from Heitler-London
 approximation. Upper graph: $\om_x$ was fixed ($\hbar \om_x=3$ meV), 
 $\om_y$ was varied; interdot distance $a$=0.7 $a_{Bx}$.
Bottom graph: $\om_y$ was fixed ($\hbar \om_y=3$ meV), $\om_x$ was varied;
interdot distance a is the same as in the upper graph.
For both graphs, solid line: magnetic field $B$=0.8T; dashed-dot line: 
$B$=1.2T; dotted line: $B$=4T. 
\bigskip

\noindent Figure 3. Exchange energy $J$ in units of meV plotted against
anisotropy $\om_y/ \om_x$, as obtained from 
Hund-Mulliken approach. Upper graph: $\om_x$ was fixed ($\hbar \om_x=3$ meV),
$\om_y$ was varied; interdot distance $a$=0.7 $a_{Bx}$. 
Bottom graph: $\om_y$ was fixed ($\hbar \om_y=3$ meV), $\om_x$ was varied;
interdot distance a is the same as in the upper graph.
Solid line: $B$=0.8 T; dashed-dot line: $B$=1.2 T; dotted line: $B$=4 T.

%\pagebreak
%\begin{figure}
%\centerline{\psfig{file=fig1.eps,width=81mm}}\bigskip
%\caption{}
%\end{figure}
%\bigskip
%\begin{figure}
%\centerline{\psfig{file=fig2.eps,width=85mm}}\bigskip
%\caption{}
%\end{figure}
%
%\bigskip
%\begin{figure}
%\centerline{\psfig{file=fig3.eps,width=85mm}}\bigskip
%\caption{}
%\end{figure}
\end{document}